\documentclass[11pt]{article}
\usepackage{authblk}
\usepackage{amssymb}
\usepackage{amsmath}
 \usepackage{graphicx}
\usepackage{color}

\def\ga{{ \gamma}}

\def\eps{{ \epsilon}}

\newcommand{\nn}{\nonumber}
\newcommand{\dis}{\displaystyle}

\newcommand{\mmmintone}[1]{{\dis{\int\kern -.36cm-}}_{\kern-.21cm\substack{#1}}\;\;}
\newcommand{\mmmintwo}[2]{{\dis{\int\kern -.43cm-}}_{\kern-.21cm\substack{#1}}^{\substack{#2}}\;\;}
\newcommand{\submint}{{\scriptstyle{\int\kern -.66em -}}}
\newcommand{\submintone}[1]{{\scriptstyle{\int\kern -.66em-}}_{\scriptscriptstyle{\kern-.21em\substack{#1}}}}
\newcommand{\fracmint}{{\textstyle{\int\kern -.88em -}}}
\newcommand{\fracmintone}[1]{{\textstyle{\int\kern -.88em
-}}_{\scriptscriptstyle{\kern-.21em\substack{#1}}}\;}

\title{Latent heat and the Fourier law}

\author{M.Colangeli \footnote{Gran Sasso Science Institute, Viale F. Crispi 7, 00167 L' Aquila, Italy.\\ E-mail: matteo.colangeli@gssi.infn.it}, A. De Masi\footnote{Universit\`{a} degli Studi dell'Aquila, Via Vetoio, 67100 L'Aquila, Italy.\\ E-mail: anna.demasi@univaq.it}, E. Presutti \footnote{Gran Sasso Science Institute, Viale F. Crispi 7, 00167 L' Aquila, Italy.\\ E-mail: errico.presutti@gmail.com}}
\date{\today}

\begin{document}

\maketitle

\begin{abstract}
\noindent
We present  computer simulations run with a  stochastic  cellular automaton
which describes $d=1$ particle
systems connected to reservoirs which keep two different densities at the endpoints.  We fix the parameters so that there is a phase transition (of the van der Waals type) and observe that if the densities at the boundaries
are metastable then, after a transient, the system reaches an apparently stationary regime where the current flows from the reservoir with smaller density to the one with larger density.\\

\end{abstract}

%\footnotetext[1] {Gran Sasso Science Institute, Viale F. Crispi 7, 00167 L' Aquila, Italy. E-mail: matteo.colangeli@gssi.infn.it.}
%\footnotetext[2] {Universit\`{a} degli Studi dell'Aquila, Via Vetoio, 67100 L'Aquila, Italy.\\ email: anna.demasi@univaq.it.}
%\footnotetext[3] {Gran Sasso Science Institute, Viale F. Crispi 7, 00167 L' Aquila, Italy.\\ e-mail: errico.presutti@gmail.com.}

\section{Introduction}
\label{sec.0}

The Fourier law states that the heat flux is proportional to minus the gradient
of the temperature, analogously the Fick law says that the mass flux is
proportional to minus the gradient of the mass density.  Both laws state that
a gradient gives rise to a current. On the other hand in the presence of a
first order phase transition there is a spontaneous separation of phases giving rise to a gradient (of the corresponding order parameter) without a current.  Purpose of this article is to investigate how this fits with  the Fourier  or the Fick law, in particular to understand the role of the latent heat in heat conduction.
In the sequel we will however refer to mass transport (hence to
the Fick law), as we will study particles models.

The physical system we have in mind is made by a channel containing a gas of particles and
by two density reservoirs
which are respectively connected to the right and to the left of the channel and which fix
the density of the gas at the endpoints
of the channel at values $\rho_+$ and, respectively,  $\rho_-$.  We further suppose that the temperature is fixed throughout
the channel at a value for which there is a phase transition.

We model the channel as one-dimensional and the gas as a system of particles which interact via a two-body attractive Kac potential, which in the Kac scaling limit gives rise to a van der Waals phase transition.  We actually consider two models, the first one (described in Section \ref{sec.2}) is a lattice gas with Kawasaki dynamics and Kac potential, the second one (described in Section \ref{sec.1}) is a stochastic cellular automaton
(CA) whose updating rules mimic the Kawasaki dynamics of the first one.  While the first model is convenient for a theoretical analysis, the second one is amenable to computer simulations.  Unfortunately, we cannot go very far theoretically and our results rely essentially on the simulations.

The  simulations exhibit two totally unexpected phenomena when the reservoirs densities
$\rho_-$ and $ \rho_+$ are such that
$\rho_- < \rho_+$, and for the gas in the channel these values are minus/plus metastable (i.e.\ metastable and in the two different phases).  In such a case
the system seems to reach a stationary state such that
(1)\;  the current in the channel becomes positive so that mass goes from the reservoir at lower density to the one with larger density; (2)\;  in a large fraction of the volume the density is metastable.  We will argue in Section \ref{sec.2} that this does not contradict the Fick law, but our arguments are not mathematically complete.
A consequence of (1) is the theoretical possibility of constructing circuits
made of the above channel connected to two large but finite reservoirs which also exchange mass with each other (either directly or via a second channel where the gas has no phase transitions).  Preliminary simulations seem to indicate that, in the circuit, after a transient,
there is a stationary current which runs in the absence of
an  external bias.  We believe that such a state is metastable with a very long life, but that in the long run the system will eventually decay to a state with no current.

%
% tw
%We will also consider the case when the density reservoirs are not infinite and they exchange mass not only with the channel but also among themselves
%so that the channel plus the two reservoirs forms a closed circuit.  The simulations indicate that after a suitable time transient the system reaches
%a stationary state  (on the times of the simulations) with
%a non zero current.
%
%In Section \ref{sec.1} we describe the cellular automaton (CA) and present the results of our computer simulations. In Section \ref{sec.2} we describe the lattice gas with Kac interaction; we explain why it is related to the cellular automaton and then give an interpretation of the simulations.  In Section \ref{sec.3} we discuss the model with finite reservoirs and give some concluding remarks.

\section{The simulations }
\label{sec.1}

Our simulations use a CA introduced in
\cite{LOP91}
to simulate the time evolution of
a system of particles which undergoes a phase transition of  van der Waals type.
The CA describes a system of particles in the interval $[1,L]$ of $\mathbb Z$,
hereafter called ``channel''. The particles
have only velocities $v\in\{-1,1\}$ and
we impose single occupancy, namely there cannot be two particles at  same site
with same velocity, $\eta(x,v) \in \{0,1\}$ being the occupation variable at $(x,v)$.

The definition of the CA involves five parameters, $L$, $\ga: \ga^{-1}\in \mathbb N$,
%$\ga^{-1}<L$
  $C>0$
and $0\le \rho_- < \rho_+ \le 1$.  We use the following notation: for $x \in [1,L]$, $\eta(x):=\eta(x,-1)+\eta(x,1)$; for $x\ge 1$,  $\eta^{(+)}(x)= \eta(x)$ if $x \in [1,L]$
and $\eta^{(+)}(x)= 2\rho_+$ if $x >L$; for $x\le L$,  $\eta^{(-)}(x)= \eta(x)$ if $x \in [1,L]$
and $\eta^{(-)}(x)= 2\rho_-$ if $x <1$; finally for $x \in [1,L]$ we call
\[
N_{+,x,\ga}
= \sum_{y=x+1}^{ x+\ga^{-1}}\eta^{(+)}(y),\;
N_{-,x,\ga}
= \sum_{y= x-\ga^{-1}}^{x-1}\eta^{(-)}(y)
\]

We are now ready to define how the CA operates.
 The unit time step updating is obtained as the result of two successive
operations: (1) {\em velocity flip}.  At all sites $x\in [1,L]$ where there is only
one particle we update the velocity of the particle to become $+1$ with probability $\frac 12
+ \eps_{x,\ga}$ and $-1$ with probability $\frac 12
- \eps_{x,\ga}$, $\eps_{x,\ga}= C\ga^2[N_{+,x,\ga}-N_{-,x,\ga}]$.
At all other sites the occupation numbers are left unchanged. Moreover, after adding
two auxiliary sites $0$ and $L+1$, we put  a particle in $0$ with velocity $+1$
with probability $\rho_-$, while we leave it empty with complementary probability; analogously we put  a particle in $L+1$ with velocity $-1$
with probability $\rho_+$ while we leave it empty with complementary probability.
(2) {\em advection}. Each particle moves by one lattice step in the direction of its velocity,
if it goes to $L+1$ or to $0$ it is deleted.

{\em Remarks.}  $\eps_{x,\ga}$ is a ``small bias'' (for $\ga$ small) which directs the velocity towards regions with higher density.  As discussed in the next section, this can be interpreted as the action of two-body  ``long range'' attractive forces; in such a context the constant $C$
is proportional to the inverse temperature $\beta$,  $2C=\beta$,  and in the limit as
$\ga\to 0$ the equilibrium phase diagram exhibits a van der Waals phase transition
for all  $C>0.5$.  The addition of the extra sites $0$ and $L+1$ in the definition of the CA simulates the action of the two reservoirs which after each time step put a new particle
at $0$ and at $L$ with probability $\rho_-$ and respectively $\rho_+$.
The action of the reservoirs is however twofold: in fact, besides the aforementioned insertion of particles in the channel with probabilities $\rho_{+}$ and $\rho_-$,
it also enters in the definition of  $\eps_{x,\ga}$, where the occupation numbers  at $y>L$ and $y<1$ are replaced by the average reservoir densities $\rho_+$ and, respectively, $\rho_-$.

We have run several  Monte Carlo  simulations for different values of the parameters defining the CA, we report here results in the case $C=1.25$, $\ga^{-1}=30$, $L=600$ and $\rho_- < \rho_+ = 1-\rho_-$.
We have computed the local particles
density $\rho(x,t)$ by taking the time average $\frac{1}{2T} \sum_{s=t}^{t+T-1} \eta_{s}(x)$, $\eta_{s}(x)$ the number of particles at $x$ at time $s$,
$T=L^2$;
however, instead of
$\rho(x,t)$ we have plotted  $ m(r,t) = 2\rho(\ga^{-1}r,t)-1$, thus the unit space length becomes $\ga^{-1}$ (the interaction range)
and the density is written in ``magnetization variables'' so that the magnetization
at the endpoints is $m_+ =-m_-$.

\begin{figure}[h]
\centering
\includegraphics[width=9cm]{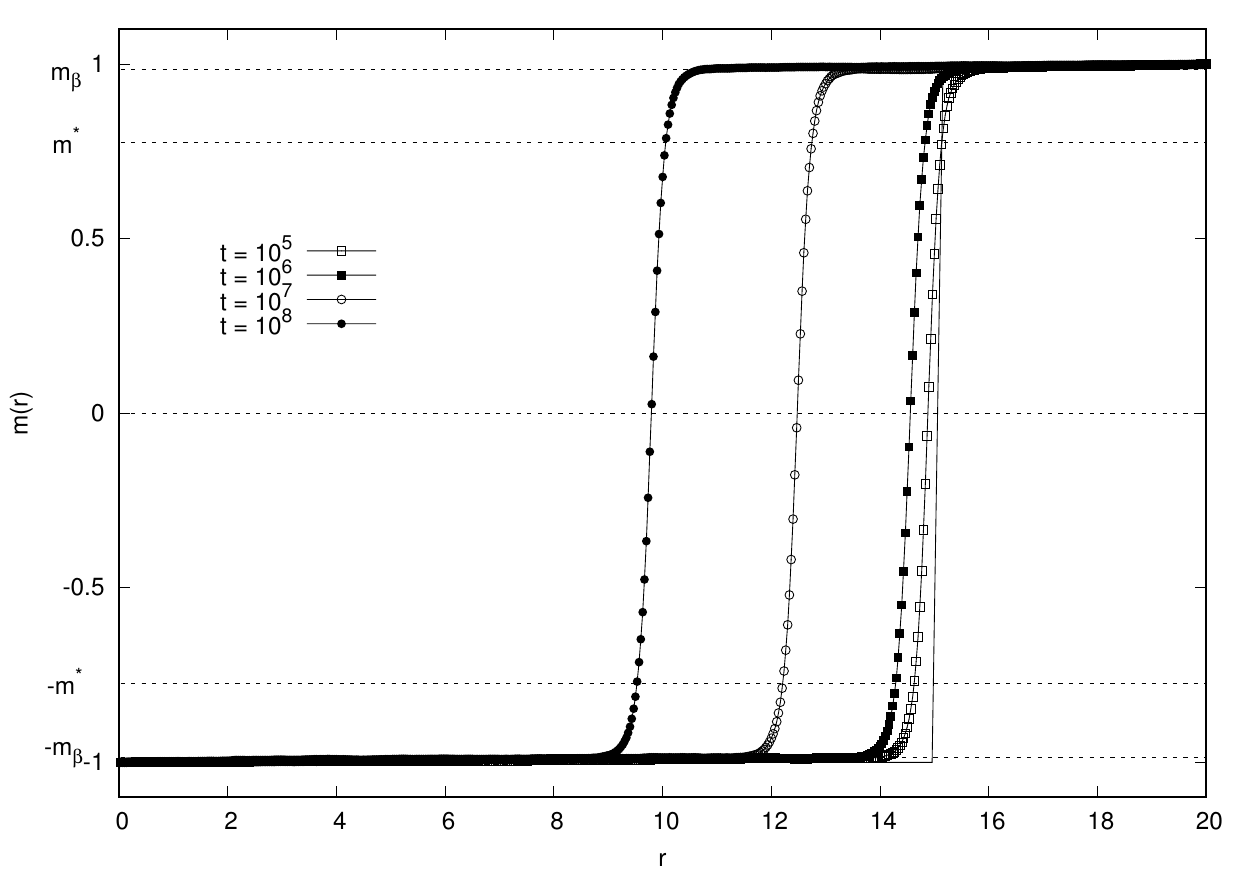}
\caption{{\footnotesize Magnetization profiles for $C=1.25$
and   $m_+ =1$ with space  in $\ga^{-1}$ ($=30$) units. The parameters $m_\beta$ and $m^{*}$ have values $m_\beta=0.985$ and $m^{*}=0.775$.  The different curves in the plot correspond to the averaged magnetization computed at different times: $t=10^5$ (empty squares), $t=10^6$ (filled squares), $t=10^7$ (empty circles) and $t=10^8$ (filled circles). The black thin line denotes the initial configuration, corresponding to a step function centered at $r=15$.}}
\label{fig:1}
\end{figure}

In Fig.~\ref{fig:1}
we report what observed when $m_{\pm}=\pm 1$ while
the initial configuration
has $m_{0,x}=-1$ for $x \le 3L/4$ and $m_{0,x}=+1$ elsewhere.
On the time scale $L^2$ we see the initial step to smoothen out: the profile becomes a
curve starting on the left at $m_-=1$ and increasing slowly, almost linearly, till $3L/4$
where it has a value $\approx -m_\beta$, $m_\beta = 0.985$, then there is
a transition region where  the magnetization
increases quite abruptly from $-m_\beta$
to $m_\beta$; afterwards the profile goes again slowly, almost linearly, up to $m_+=1$ which is reached at the right endpoint.  As time increases the profile  moves rigidly towards
the middle of the channel which is reached on times $\le L^3$ and in the time of our simulations it
remains unchanged except for small fluctuations. In the next section we
will interpret the values $\pm m_\beta$ as the equilibrium magnetization densities
when the inverse temperature is
$\beta = 2C$.

\begin{figure}[h]
\centering
\includegraphics[width=9cm]{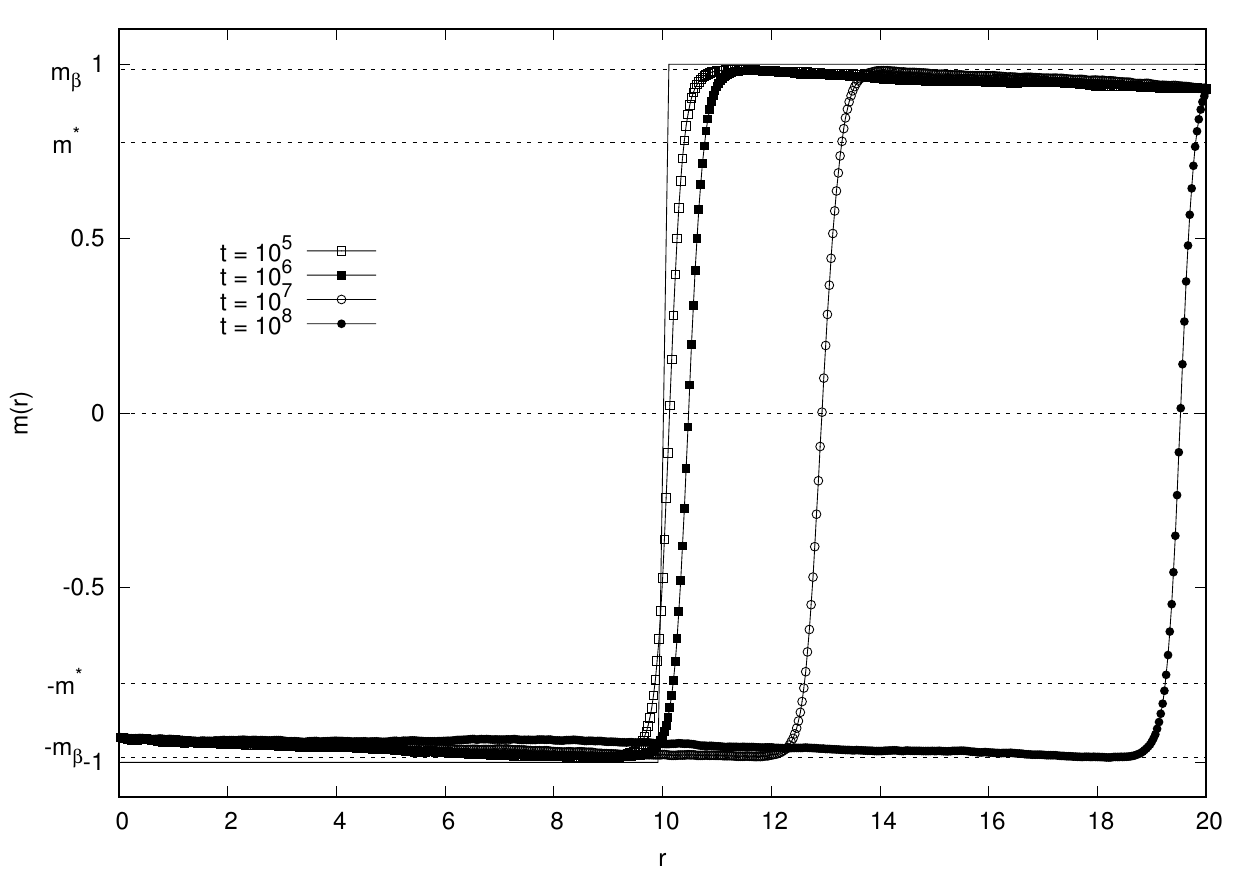}
\caption{{\footnotesize Magnetization profiles for $C=1.25$, $m_\beta=0.985$ and $m^{*}=0.775$, and with  $m_+ =0.93$. The curves in the plot have the same meaning of those illustrated in Fig.~\ref{fig:1}. The initial datum is a step function centered at $r=10$.}}
\label{fig:2}
\end{figure}

If we decrease   $m_+$ till $m_\beta$
we see the same pattern with a transition region which is essentially unchanged and the quasi linear parts with a smaller slope.
However if $m_+$ decreases past   $ m_\beta$ keeping $m_+ > m^{*}=0.775$
we see  a completely different picture (as argued in the next section, the values $|m| \le m^*$ are to be regarded as unstable, $m^*< |m| < m_\beta$ as metastable, and $|m| \ge m^*$ as stable). In Fig.~\ref{fig:2}
we report simulations with  $m_+=0.93$.
We start now from an initial configuration
which has $m_{0,x}=-1$ for $x \le  L/2$ and $m_{0,x}=+1$ elsewhere.
We observe, after a short transient, a pattern similar to the one in Fig.~\ref{fig:1},  i.e.\  with a transition region around the middle which is very similar to the previous one.
To its right and left there are again approximately linear profiles but now they are
decreasing (because $m_+ < m_\beta$). In contrast to the previous case as time increases on the scale
$L^2$   the transition region moves away from the middle and on times
$L^3$ it ``collides'' with an endpoint of the channel:
in Fig.~\ref{fig:2} it is represented by a  bump on the right of the channel where the magnetization rapidly increases from $-m_\beta$ to $m_+$, $m_+$  the magnetization forced by the right reservoir. If we change the seed of the random generator we may as well see the bump on the left.
Such a profile seems stationary as it stays unchanged (modulo small fluctuations) for
very long times, our longest simulation has $t=10^{11}$.

\begin{figure}[h]
\centering
\includegraphics[width=9cm]{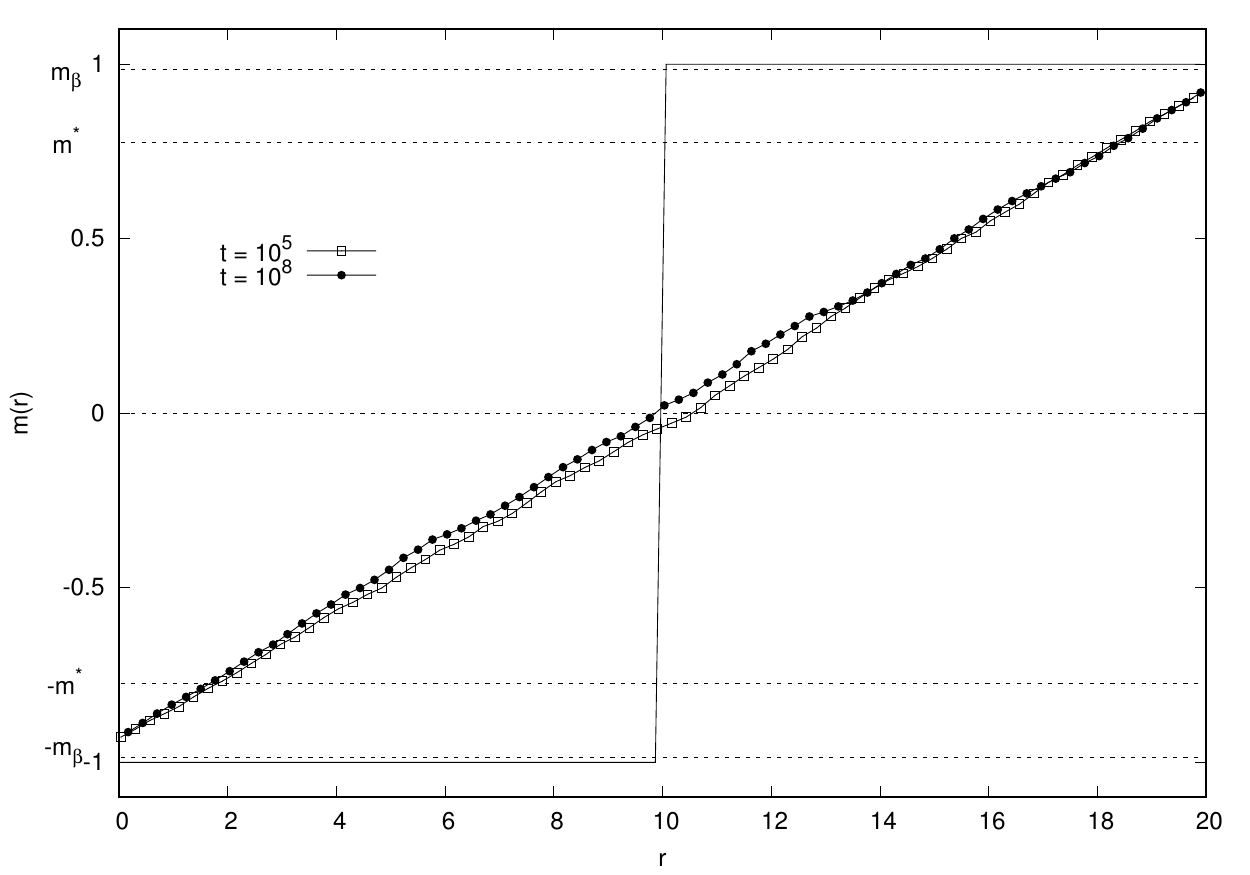}
\caption{{\footnotesize Magnetization profiles for $C=0$, $m_\beta=0.985$ and $m^{*}=0.775$, and with  $m_+ =0.93$. The curves in the plot have the same meaning as in Fig.~\ref{fig:1}.  The initial datum is a step function centered at $r=10$. The standard Fourier law is satisfied.}}
\label{fig:3}
\end{figure}

Besides the magnetization profiles, we have also
measured the current by summing (with sign) the total number of particles
which in a time interval $T$ at each time step enter into the system from site 0 to site 1
minus those which exit from the channel going from site 1 to site 0.  As the current is small, of the order $10^{-5}$, to have reliable values we had to use
longer time averages,  $T \approx L^3$.
For values $m_+>m_\beta$   the CA reaches a stationary pattern (on the time scale of our simulations) with a negative current (flowing from $L$ toward $1$) which is proportional to $1/L$.  The magnetization is increasing and essentially linear away from the short transition region and, in agreement with the Fick's law, the current is negative being proportional to minus the magnetization gradient, see the next section for more details.  Without bias (i.e.\ $C=0$) the slope would be all the way linear and therefore the current larger,  as shown in Fig.~\ref{fig:3} (when $C=0$ the values $m_\beta$ and $m^{*}$ do not play any role and the thick black dashed lines corresponding to $\pm m_\beta$ and $\pm m^*$ are drawn only to permit an easier comparison with Fig.~\ref{fig:2}).
Thus the effect of the latent heat, responsible for the occurrence of the transition region, is to reduce the conductivity of the system.

The effect becomes dramatic once we reach values of $m_+$ as those illustrated in Fig.~\ref{fig:2}.  In this regime the
current becomes positive, it goes from the left where the density is smaller to the right where the density is larger.  The profile is essentially linear away from the bump but it is now decreasing, hence the change in direction of the current. The Fick's law is still satisfied as the current goes opposite to the gradient (except in the transition region which however occupies a small fraction of the volume) and the overall effect is that the reservoir with smaller density gives mass to the one with larger density.  Same phenomenon is observed for other initial conditions and /or different seeds of the random generator.

\section{Mesoscopic limit}
\label{sec.2}

To interpret the above simulations %in a statistical mechanics framework
we will relate our CA to particle models used in non-equilibrium statistical mechanics.
The connection comes by studying the mesoscopic limit (described below) of the CA which we argue
(but do not have yet a proof) to be
the same as that obtained in the same limit
from  a particle system with Kawasaki dynamics involving Kac  potentials.

The mesoscopic limit describes the evolution of the system
in the limit  $\ga\to 0$ when space is scaled by $\ga^{-1}$
and time by $\ga^{-2}$.  Thus  $L= \ga^{-1}\ell$ with $\ell >0$
fixed independently of $\ga$, $\ell$ is the mesoscopic length
of the channel.  Let $\eta(x,v;t)$ be
the occupation number at $(x,v)$ and at time $t$
and call $u_\ga(x,v;t) =  E[\eta(x,v;t)]$ the average
particle number at $x,v;t$. We conjecture that
for a suitable choice of the initial configurations, for any $r \in (0,\ell)$, $t>0$
and $v\in \{-1,1\}$
\begin{equation}
\label{2n.1.1}
\lim_{\ga\to 0, \ga x \to r}u_\ga(x,v;\ga^{-2}t) = \rho(r,t)
\end{equation}
with $\rho(r,t)$ a continuous function with limits $\rho_{\pm}$
as $r \to \ell$ and, respectively, $r\to 0$, which satisfies (in a weak sense) the integro-differential equation

%This is, $L= \ga^{-1}\ell$, $\ell >0$, time is scaled by $\ga^{-2}$,
%.
%Let
%\begin{equation}
%\label{2n.1}
%u_\ga(x,v;t) =  E[\eta(x,v;t)]
%%,\quad
%%u_\ga(x;t) =  \frac 12\Big(u_\ga(x,-1;t) + u_\ga(x,1;t)\Big)
%\end{equation}
%where  $\eta(x,v;t)$ is the occupation number at time $t$ and at $(x,v)$; $E$ denotes  expectation.
%
%Suppose there is a non negative function $\rho(r,t)$, $r\in[0,\ell]$, $t\in [0,T]$, $T>0$, such that for all $r\in (0,\ell), t\in [0,T], v \in\{-1,1\},\delta>0$
%\begin{eqnarray}
%\label{2n.1.1}
%&&\lim_{\ga\to 0, \ga x \to r}u_\ga(x,v;\ga^{-2}t) = \rho(r,t)\nn
%%,\quad
%%r\in (0,\ell), t\in [0,T], v \in\{-1,1\}
%%\end{equation}
%%We also suppose that
%%\begin{equation}
%%\label{2n.1.2}
%\\&&\lim_{\ga\to 0, \ga x \to r}E[\eta(x,v,\ga^{-2}t)\{1-
%\eta(x,-v,\ga^{-2}t)\}] = \rho(r,t)\{1-\rho(r,t)\} \nn
%%,\quad
%%r\in (0,\ell), t\in [0,T]
%%\end{equation}
%%\begin{equation}
%%\label{2n.1.3}
%\\&&\lim_{\ga\to 0, \ga x \to r}P[|\eps_{\ga,x,\ga^{-2}t}-
%\int_{r}^{r+1} d\xi[\rho(r+\xi,t)
% - \rho(r-\xi,t)]|>\delta] = 0
% %,  t\in [0,T], v \in\{-1,1\}, \delta>0
%\end{eqnarray}
%Then  $\rho(r,t)$ satisfies
\begin{equation}
\label{2n.1.4}
 \frac{\partial}{\partial t}\rho(r,t) = \frac 12 \frac{\partial^2}{\partial r^2}\rho(r,t)
 - 4C \frac{\partial}{\partial r}\Big(\rho(r,t)[1-
   \rho(r,t)]\int_{r}^{r+1} d\xi[\rho(r+\xi,t)
 - \rho(r-\xi,t)]\Big)
 \end{equation}
 where $\rho(r,t)= \rho_{\pm}$ if $r\ge \ell$ and respectively $r\le 0$.
%weakly against   test functions $f(r,t)$ with compact support in $(0,\ell)\times (0,T)$.  We need also to suppose that $\rho(r,t)$ satisfies the boundary conditions
%\begin{equation}
%\label{2n.1.5}
%\rho(0,t)=\rho_-, \; \rho(\ell,t)=\rho_+
%\end{equation}
%so that \eqref{2n.1.4}--\eqref{2n.1.5} specify $\rho(r,t)$.
The above statements can be proved under the assumption that
``propagation of chaos'' holds in some strong form, a real proof
is in preparation.

It is now convenient to switch to spin variables, so we define
\begin{equation}
\label{2n.10}
m(r,t) = 2\rho(r,t)-1;\; \rho(r,t)= \frac {m(r,t)+1}2
 \end{equation}
Then \eqref{2n.1.4} becomes
\begin{equation}
\label{2n.11}
 \frac{\partial}{\partial t}m(r,t) = \frac 12 \frac{\partial^2}{\partial r^2}m(r,t)
 - C \frac{\partial}{\partial r}\Big([1-
 m(r,t)^2]\int_{r}^{r+1} d\xi[m(r+\xi,t)
 - m(r-\xi,t)]\Big)
 \end{equation}
which, setting $\beta = 2C$,  can be rewritten as
\begin{equation}
\label{2n.12}
 \frac{\partial}{\partial t}m  = \frac 12\frac{\partial}{\partial r}\Big(
 \frac{\partial m}{\partial r}
 - \beta [1-
 m^2]\int_{r}^{r+1} d\xi[m(r+\xi,t)
 - m(r-\xi,t)]\Big)
 \end{equation}
\eqref{2n.12} is the conservation law
\begin{equation}
\label{2n.13}
\frac{\partial m}{\partial t}=  -\frac {\partial }{\partial r}   I
\end{equation}
where
\begin{eqnarray}
\label{2n.14}
 &&
 I(r)= - \chi \frac {\partial}{\partial r}\frac {\delta F(m)}{\delta m(r)},\quad \chi =\frac\beta 2(1-m^2)
 \\&& F(m) = \int dr \Big(-\frac {m^2}{2} - \frac {S}{\beta}\Big) + \frac 14\int dr\int dr'
 J(r,r') [m(r)-m(r')]^2
 \nn
 \\&&S(m)= -\frac{1-m} {2}\log \frac{1-m} {2} - \frac{1+m} {2}\log \frac{1+m} {2}
\nn
\end{eqnarray}
with $J(r,r') = 1-|r-r'|$ for $|r-r'|\le 1$ and $=0$ elsewhere.

\eqref{2n.13}--\eqref{2n.14} have been derived in \cite{GL97} as the mesoscopic limit
of an Ising model with Kawasaki dynamics and Kac potential
$J_{\ga}(x,y) = \ga J(\ga x,\ga,y)$  (in \cite{GL97} the system is in a torus,
the case $d>1$ is also covered).  Thus the interaction term $\eps_{x,\ga}$
in the simulations can be regarded as a force due to an attractive pair Kac potential.

\begin{figure}[h!]
\centering
\fbox{\includegraphics[width=6.5cm]{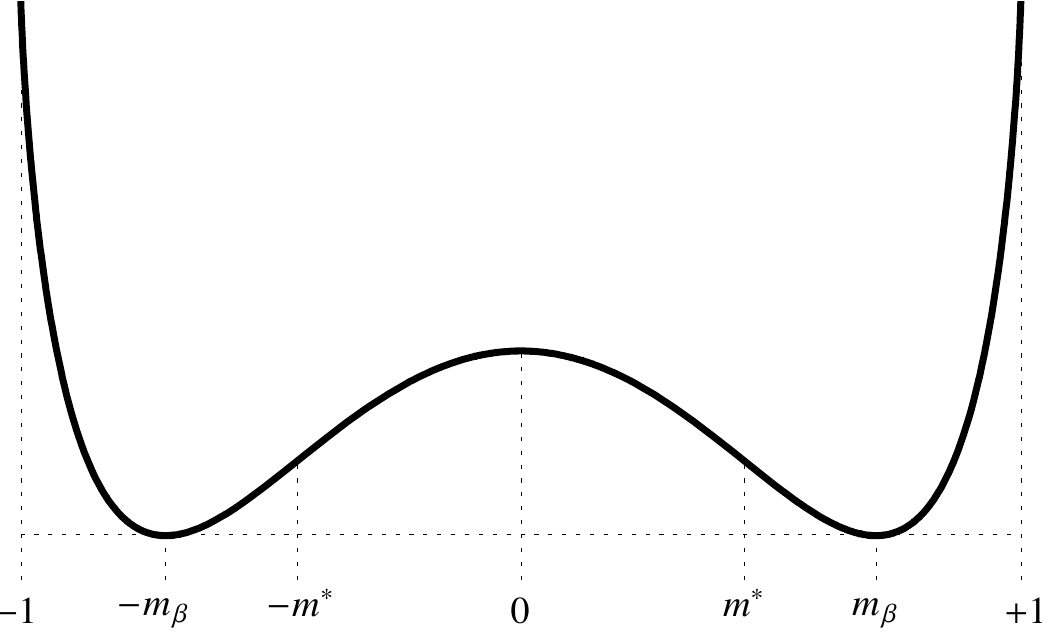}}
\caption{{\footnotesize The graph of the excess free energy $f_\beta(m)-f_\beta(m_\beta)$, see Eq. \eqref{2n.14}, as a function of $m$, for $\beta= 1.25>\beta_c=1$.}}
\label{fig:5}
\end{figure}

The Ginzburg-Landau free energy functional can be seen as a local
approximation of $F(m)$ when $J(x,y)$ becomes a delta function,
so that the non local term in $F(m)$ becomes a gradient squared.  Correspondingly, the conservative gradient flow for  Ginzburg-Landau, which is the Cahn-Hilliard equation, is a local approximation of the conservative gradient flow for $F(m)$ which is  \eqref{2n.13}--\eqref{2n.14} and in this sense our CA simulates the Cahn-Hilliard equation.  The van der Waals free energy associated to $F(m)$ is
\begin{eqnarray}
\label{1.14}
 &&
 f_\beta(m)=  -\frac {m^2}{2} - \frac {S}{\beta}
\end{eqnarray}
see Fig.~\ref{fig:5}.

For $\beta>1$ this is a double well potential with minima at $\pm m_\beta$ where
$m_\beta$ is the positive root of the mean field equation
\begin{eqnarray}
\label{1.15}
 &&
  m_\beta =  \tanh\{\beta m_\beta\}
\end{eqnarray}
The van der Waals- Maxwell free energy is the convex envelope
$f^{**}_\beta(m)$ of $f_\beta(m)$, which is constantly equal to $f_\beta(m_\beta)$ in the interval $[-m_\beta,m_\beta]$ and it is equal to $f_\beta(m)$ elsewhere.
The values $|m| \ge m_\beta$ are stable.  The values $|m| \in (m^*,m_\beta)$ are metastable,
\begin{eqnarray}
\label{1.16}
 &&
  m^*>0 :  \beta(1-(m^*)^2) =1
\end{eqnarray}
while the values $|m| \le m^* $ are unstable.  In the stable and metastable regions $f_\beta(m)$ is convex, in the unstable region it is concave.

Let us now go back to \eqref{2n.14} and
given a magnetization profile $m(r)$ define the molecular magnetic field $h(r)$ as
\begin{eqnarray}
\label{1.17}
 &&
 h(r)=  \frac {\delta F(m)}{\delta m(r)}
\end{eqnarray}
%$h(r)$ is the magnetic field which produces the profile $m$ in the sense that given $h(r)$, $m(r)$ is the solution of the non local mean field equation
%\begin{eqnarray}
%\label{1.18}
% &&
% m(r) = \tanh\Big\{\beta [\int J(r,r')m(r')dr' + h(r)]\Big\}
%\end{eqnarray}
With these notation  \eqref{2n.14} reads as
\begin{eqnarray}
\label{1.19}
 &&
I(r)= - \chi \frac {\partial h(r)}{\partial r}
\end{eqnarray}
which looks like the Fick law for the magnetization current written with the gradient of the magnetic field rather than the gradient of the magnetization density.  However the relation between the
magnetic field $h(r)$ and the magnetization density $m(r)$ is not given by the thermodynamic relation between magnetization and magnetic field, in particular it is non
local.  To gain locality and the correct relation
we should go from the mesoscopic to the macroscopic description of the system
in the limit when $\ell \to \infty$ and with space rescaled by a factor $\ell^{-1}$.

The simulations seem to indicate that in this limit the mesoscopic profile (in the
context of Fig.~\ref{fig:2}) becomes a smooth almost linear decreasing profile starting
from the left endpoint where it has  value $m_-$  and converging to $-m_\beta$ at the right endpoint, here there is a boundary layer effect as the bump shrinks to a point in the macroscopic variables.
Thus
except for a set of Lebesgue measure 0 (the point to which the transition region shrinks) the profile is monotone (decreasing) and the rescaled current becomes equal  to
$-  \frac{\partial f_\beta(m)}{\partial m}\frac{\partial m(r)}{\partial r}$ so that the Fick law is satisfied.  However, the limit profile $m(r)$ is all contained within the metastable region so that
we do not get the true thermodynamic relation which instead involves $f^{**}_\beta(m)$.
This is the other main unexpected phenomenon which emerges from our simulations.

Summarizing, the patterns as in Fig.~\ref{fig:1} appear when $m_+ > m_\beta$, those in Fig.~\ref{fig:2} when
$m_+\in (m^*,m_\beta)$.  The magnetization in the transition region of Fig.~\ref{fig:1} goes from $\approx -m_\beta$ to $\approx +m_\beta$; in the bump of Fig.~\ref{fig:2} the magnetization goes from $\approx -m_\beta$ to $m_+$.  In \cite{MPT11} it is proved that \eqref{2n.12}  has (for $\ell$ large enough) a stationary solution as in   Fig.~\ref{fig:1}; it is also shown that when
$m_+\in (m^*,m_\beta)$ there is a stationary solution with the transition region
in the middle and positive current, but our simulations show that such
a profile is unstable so that the CA does not see it except in a transient.  We conjecture that arguments similar to those used in  \cite{MPT11} could prove the existence of stationary solutions with a bump.  Such a result
is proved in \cite{DOP98}--\cite{DOP00} in the semi-infinite case where $\ell \to \infty$ and using Neumann reflecting conditions in the definition of the Kac interaction.

We have an indirect proof of the validity of our conjecture on the mesoscopic limit via
the
measure of the limit current $j$.  Recall that we have  measured the current by taking the time average of the number of particles
which at each time step jump from 0 into the system minus those which exit from  the channel by jumping to 0. In the context of   Fig.~\ref{fig:2} and taking the averaging time $T$ equal to $T=L^3$, we have obtained $j = 3.97 \times 10^{-5}$.
Supposing the profile $m(r)$ stationary then $j$ should be
close to $\ga I$ (the factor $\ga$ is due to the change of scales in the mesoscopic limit) where $I$ is as in \eqref{2n.13}--\eqref{2n.14}
and computed using  $m(r)$. One can check that
% Assume that there are $r_1$ and $r_2$ with $r_1< r_2-4$ such that
if $m(r)$ is a stationary solution of \eqref{2n.12} in $[0,\ell]$ and
there is an interval $[r_1,r_2]$, $r_2-r_1>4$ where $m$ is negative and such that  $\frac {\partial m(r)}{\partial r}\le 0$  and $\beta (1-m(r)^2)\le 1$ for $r\in [r_1,r_2]$, then $I \in [I_-,I_+]$
   where
   \begin{eqnarray*}
&&I_+ = \frac{1}4\frac 1{r_2-r_1-2}\Big\{[m(r_1+1)-m(r_2-1)] \\&&-\beta(1-
 m(r_2-1)^2)[m(r_1+2)-m(r_2-2)]\Big\}\\
 %	\end{eqnarray*}
% 	\begin{eqnarray*}
&&I_-=  \frac{1}4\frac 1{r_2-r_1-2}\Big\{[m(r_1+1)-m(r_2-1)] \\&&-\beta
 (1-m(r_1+1)^2)[m(r_1)-m(r_2)]\Big\}
 	\end{eqnarray*}
 According to the simulation such an interval could be  $r_1 = 0$, $r_2=18$.  We have
$m[0] = -0.93007267$; $m[1] = -0.93231821$; $m[2] = -0.93538058$;
$m[16] = -0.97950798$; $m[17] = -0.98214394$; $m[18] = -0.98471862$; with these values $\ga I_{-} = 3.32  \times 10^{-5}$ and $\ga I_{+} = 4.78 \times 10^{-5}$ while
the current computed in the simulation is $j = 3.97 \times 10^{-5}$.

{\bf Conclusions.} Our simulations seem to indicate  that  in systems which undergo a phase transition of van der Waals type we may see in the heat conduction experiment for a long time
a current flowing from the  reservoir with  smaller order parameter to the one with the higher value and thus construct machines as described at the end of the introduction.
%schematized in Fig.~\ref{fig:4}.

{\bf Acknowledgments.}  We are indebted to Maria Eulalia Vares and Deepak Dhar for many helpful discussions and comments.

\end{document}